# Information Requirements for Enterprise Systems

Ian Sommerville[1], Russell Lock[2] and Tim Storer[3]

[1] School of Computer Science, University of St Andrews, St Andrews, Scotland
[2] Department of Computer Science, Loughborough University, Leics., England
[3] School of Computing Science, Glasgow University, Glasgow, Scotland
ian.sommerville@st-andrews.ac.uk, r.lock@lboro.ac.uk, timothy.storer@glasgow.ac.uk

**Abstract.** In this paper, we discuss an approach to system requirements engineering, which is based on using models of the responsibilities assigned to agents in a multi-agency system of systems. The responsibility models serve as a basis for identifying the stakeholders that should be considered in establishing the requirements and provide a basis for a structured approach, described here, for information requirements elicitation. We illustrate this approach using a case study drawn from civil emergency management.

**Keywords:** requirements engineering, requirements, enterprise systems, responsibility modeling, socio-technical systems.

## 1  Introduction

The derivation of requirements for complex systems has been recognized as a major problem in industry. The system requirements are a definition of what is expected of the system. They inform the system implementation and, in some cases, serve as a basis for a contract between a system procurer and a system provider. Historically, requirements have been expressed as statements of natural language text that have set out the functionality of the system that is expected. Modern agile methods have rejected the notion of requirements as descriptions of functionality and use approaches such as user stories to describe what is expected. However, these approaches are still primarily concerned with what the system should *do*.

Behavioural approaches to requirements engineering are appropriate when systems are to be developed from scratch. However, in most organization, new systems are now created by integrating functionality from existing systems and components. In such cases, it makes little sense to specify requirements in terms of what the system should do – the functionality is already defined in these systems. Rather, we argue that it is more appropriate to consider the system requirements from an informational perspective – what information should the system provide and who needs that information to do their job.

The derivation of requirements involves extensive discussions and consultations with system stakeholders – people who may be system users, their managers or who are influenced in some way by the system. An enduring problem in requirements engineering has been how to identify the stakeholders to be consulted and how to help them articulate their requirements for a system [1]. Requirements engineering methods such as Volere [2], say little about this problem – they highlight the importance of stakeholder consultation but their only guidance of stakeholder identification is to provide a list of stakeholder types. The problems of stakeholder identification are exacerbated in situations where the system to be developed spans several organizations and these stakeholders are distributed across these organizations.

To address this problem, we have developed the notion of responsibility modeling. We explicitly identify the responsibilities of organizational stakeholders in a problem setting and draw up a model showing these responsibilities and their assignment to agents. This then serves as a basis for both identifying stakeholders and for identifying whether or not there are inconsistencies in responsibility perception in the different organizations involved.

Once stakeholders have been identified, we can then enter into discussions with them about how they do their job and what information they require to do so. The responsibility model, along with a set of standard questions, is used to facilitate that discussion and to help the requirements engineer tease out the interactions between stakeholder responsibilities. This then leads to a statement of 'information requirements' which are then used to inform the system design and implementation.

In the remainder of the paper, we discuss enterprise systems and how these are typically created by composing and configuring existing software systems or components. We go on to explain why we think information requirements are the most important type of requirement for enterprise systems and follow this with an introduction to responsibility modeling. We explain how responsibility models are used to derive information requirements and illustrate our approach with a case study of an emergency management system. We conclude with a discussion of related work and our thoughts on how this work can be taken forward.

## 2  Enterprise systems

The focus of our work for a number of years has been *enterprise systems* [3]. This term is widely used and is sometimes used synonymously with the term *ERP or enterprise resource planning systems*. Whilst ERP systems are certainly enterprise systems, we actually use the term more widely to denote systems that have the following characteristics:

1. They are multifunctional systems in that they deliver different classes of functionality. For example, an enterprise system may deliver functionality to support both sales and purchasing functions in an organization.
2. They are often oriented around one or more shared databases. The sharing of data means that data is not replicated in the organization and there are opportunities for data sharing across the different functions delivered by the system.
3. The different components of the system are self-contained systems so that they can operate with or without other components. An enterprise system may therefore be considered as a system of systems.

4. They are used by different classes of stakeholder who have different jobs in the organization. Users may have different levels of power and authority in the organization and different levels of technical expertise. The user base for these systems is therefore heterogeneous and drawn from different levels in the organization.
5. The system will have emergent behavior that cannot be predicted by an analysis of the system components. This behavior may be desirable or undesirable and is a consequence of the complexity of the relationships between the different systems in the enterprise system.

ERP systems, such as those marketed by SAP and Oracle, are enterprise systems where all of the system components are supplied by a single supplier. These ERP systems normally have a preferred mode of use and organizations that wish to use an ERP system are advised to adapt their processes to this mode of use. Typically, a single ERP system will replace a number of separate systems in an organization.

More generally, enterprise systems may include systems from a number of different suppliers. These may communicate through a shared database but may also maintain their own databases. Some component systems may be legacy systems – older systems based on obsolete technology that have been 'wrapped' with e.g. a service interface so that they can interact with other systems. Other components may be off-the-shelf systems from different manufacturers, specially written systems, etc.

Enterprise systems may be considered as technical software/hardware systems but they are an integral part of wider socio-technical systems in the enterprise. Socio-technical systems are systems that include people as well as technical elements and which are profoundly influenced by organizational policies, processes and culture, as well as external regulation. In essence, socio-technical systems are the ways in which work gets done in an organization.

Over the past decade of so, the notion of a virtual organization or virtual enterprise [4] has been developed. A virtual organization is temporary entity that is created with a particular mission and which involves a number of other organizations. For example, a virtual organization may be created to organize a major sporting event such as the Olympic Games. This encompasses many different partners, who each have their own IT systems.

Virtual organizations are enterprises in their own right and enterprise systems may therefore be created to support their operation. In this case, the component systems are distributed across the organizations in the virtual enterprise. These systems have all of the above enterprise system characteristics but with additional complications:
1. The system components in the system of systems are independently owned and managed. This means that there is no single authority that can control the functionality and development of the enterprise system.
2. There is no single shared database but rather a confederation of databases from the different organizations that are involved in the system. Inevitably there are syntactic and semantic incompatibilities between these systems.
3. The practices and cultures of the different organizations in the virtual organization are different. This has the consequence that the overall virtual enterprise system is perceived in quite different ways by stakeholders in these different organizations.

In this paper, we will draw on our experience of interacting with a virtual organization, which is created to deal with serious civil emergencies such as a terrorist attack, regional flooding, or a nuclear incident.

## 3   Requirements engineering for enterprise systems

Requirements engineering (RE) is the process of understanding a system's environment with a view to deriving the requirements for the system – what has to be implemented to provide the business functionality that is required. For organizational systems, this inevitably means dealing with multiple stakeholders from different parts of the organization who have differing needs and priorities. The RE process therefore inevitably involves negotiations with and between these stakeholders to arrive at a set of requirements that is acceptable to all stakeholders. This is illustrated in Figure 1.

The requirements engineering team works with stakeholders to understand their requirements for a new or replacement system. These requirements are then documented, usually using natural language text, and system models of different kinds may be produced. These are then taken back to stakeholders for checking. Typically, this is an incremental process and there will be several rounds of the cycle completed before a comprehensive set of requirements have been produced.

Inevitably, there will be conflicts between these requirements as they will represent the wishes of stakeholders with diverse needs. Some of these conflicts will be resolved by the requirements engineering team but there is always a need for a period of negotiation to settle disagreements and to arrive at a set of compromise requirements. This negotiation may also involve the implementation team who provide information about the costs of implementing the requirements – if requirements are too expensive to implement, they may be discarded.

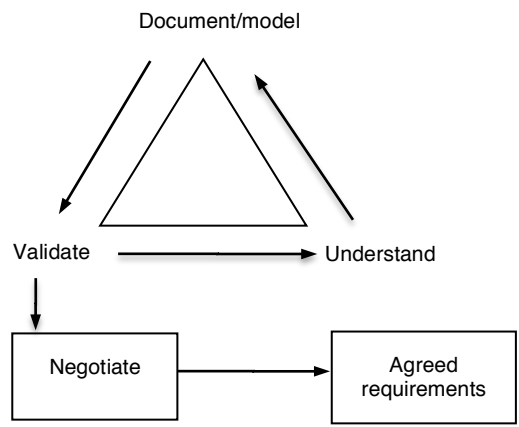

**Figure 1.** The requirements engineering cycle.

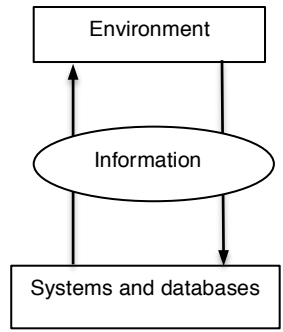

**Figure 2.** Enterprise system requirements engineering

Most approaches to requirements engineering that have been developed have adopted a behavioural perspective – they focus on what the system should 'do', in terms of delivering functionality to stakeholders of different kinds. However, when we are considering enterprise systems, the detailed functionality is largely pre-defined by the system components that are used. Instead, we argue that the focus of requirements engineering process should be on identifying the information that is needed and used by stakeholders, rather than the specific functionality that is used.

In essence, the requirements engineering process should focus its analysis on the information that people need to do their work, the information that they create in the course of that work and the information that is shared with other people. Non-functional information requirements such as confidentiality requirements for shared information, presentation requirements, etc. may also be elicited .

The basis of this idea is not simply that there is limited scope for extending the functionality of the system. It reflects the reality that the introduction of an enterprise system normally requires changes to the business processes in the organization. People have to change and adapt to use the new system and, by and large, this is not really a problem. They can learn new processes and user interfaces. Problems arise, however, when people do not have the information they need to do their job, whatever the specific process that is used.

Therefore, a focus on the information needs of stakeholders, as illustrated in Figure 2, is likely to be the most productive approach for enterprise system requirements engineering.

When we are considering information requirements, however, we need to take into account that political and organizational considerations affect both the availability and the sharing of information. Stakeholders may deliberately withhold or delay information because they see some personal benefit in doing so; they may demand that information be presented in certain ways or may insist on their own information classification schemes.

To illustrate what we mean here, consider the situation in hospitals where there is perennial (and probably inevitable) tension between the hospital administrators and the senior doctors. Information that is required to support administration is inevitably different from clinical information and providing that information often requires doctors to do extra work. If doctors are in a strong position within the organization, they may simply refuse to provide that information, thus constraining the information system. On the other hand, if the power struggle favours the hospital managers, then the doctors may comply with the demands to change the way they capture patient information. The information requirements depend on the power relationships in the organization as well as what people need in order to do their job.

## 4  Responsibility and responsibility modelling

Our work over the past few years has been concerned with the notion of socio-technical systems engineering, where we are exploring how methods and techniques for socio-technical analysis of organizations can be used alongside systems and software engineering methods [5]. As part of this work, we have been investigating the abstractions that can be used to model complex socio-technical systems. Such systems include human and automated components, are significantly influenced by organizational policies, culture and politics and often involve participants and systems from a number of different organizations.

One abstraction that we have found to be particularly helpful is the notion of 'responsibility', which can be used to represent the expectations placed on both individuals and systems and which is a universal abstraction, used in all types of organization. We define a responsibility to be:

*A duty, held by some agent, to achieve, maintain or avoid some given state, subject to conformance with organizational, social and cultural norms.*

The key points in this definition are
- a responsibility is a duty, which implies that the agent holding the responsibility is accountable to some authority for their actions,
- responsibilities may be concerned with avoiding undesirable situations and not just with accomplishing some actions
- in discharging responsibilities, agent behaviour is constrained by laws, regulations and social/cultural conventions and expectations. Therefore, the effectiveness of an agent in discharging their responsibility is not only judged by the outcome but also by the ways in which the agent has discharged that responsibility.

Responsibilities are a particularly helpful abstraction because they are firmly rooted in the world of work and are not abstract notions, such as goals, which are apparently internalized in individuals. The naturalness of responsibilities means that responsibility holders find it easy to communicate with people about their own responsibilities and also about the responsibilities of others.

Of course, it is often the case that there are different interpretations about what a responsibility means. Perceived differences in what a responsibility entails are often helpful in identifying sources of misunderstanding and, sometimes, requirement conflicts. For example, a responsibility to arrange seminars in a university may be interpreted as simply involving finding speakers and gaining their agreement to speak, but without any involvement in booking rooms, arranging refreshments, etc. The same responsibility may also be considered to be more inclusive so that it involves both finding speakers and making all other arrangements for the seminar presentation.

A responsibility model is a succinct description of the responsibilities that have been assigned to agents in one or more organizations. Our experience in modeling with client organizations is that modeling notations have to be simple, easy to explain and must avoid technical concepts that are alien to the people in the organization. For this reason, we believe that technical modeling notations such as the UML are not particularly useful for early-stage requirements engineering.

To make the models as simple as possible to explain, we have limited a responsibility model to three abstractions:

1. Responsibilities, as discussed above. Examples of responsibilities, drawn from an emergency response system, might be 'Establish local communications', 'Casualty evacuation' and 'Press liaison'.
2. Agents, which are organizational, human or system entities that may be assigned responsibilities. Therefore, an agent may be a named organization such as the ambulance service, a person or a role, such as the communication coordinator or a software-intensive system, such as an automated despatcher for emergency vehicles.
3. Resources, which are used by agents in discharging their responsibilities. We distinguish between two types of resource namely physical resources, which are 'consumed' in use and information resources, which are not. An example of a physical resource is an ambulance – there are a limited number of ambulances in an area and once these have all been allocated, the despatcher must wait until one has been released. By contrast, an information resource such as a geographical information system is not (normally) limited by demand – it can be used irrespective of the number of users.

Figure 3 illustrates a responsibility model that we developed as part of an analysis of response to a civil nuclear emergency at a power station by the coast. There are consequent responsibilities to inform shipping in the area. In this model:

1. Responsibilities are shown in round-edged rectangles.
2. Agent names are enclosed in angle brackets.
3. Physical resources are shown in square brackets.
4. The names of information resources are surrounded by vertical bars.
5. Arrows show the sources and destination of information.

From this model, you can see that responsibility to check on the safety of shipping falls on MRCC Clyde, the maritime rescue coordination center for the Clyde Estuary area and it relies on incident information provided by the police and the nuclear emergency liaison officer from the NAECC, the National Atomic Emergency Coordination Centre. Notice that we don't decompose this responsibility – how it is discharged is up to the organization assigned the responsibility and is of no concern to the emergency coordination team.

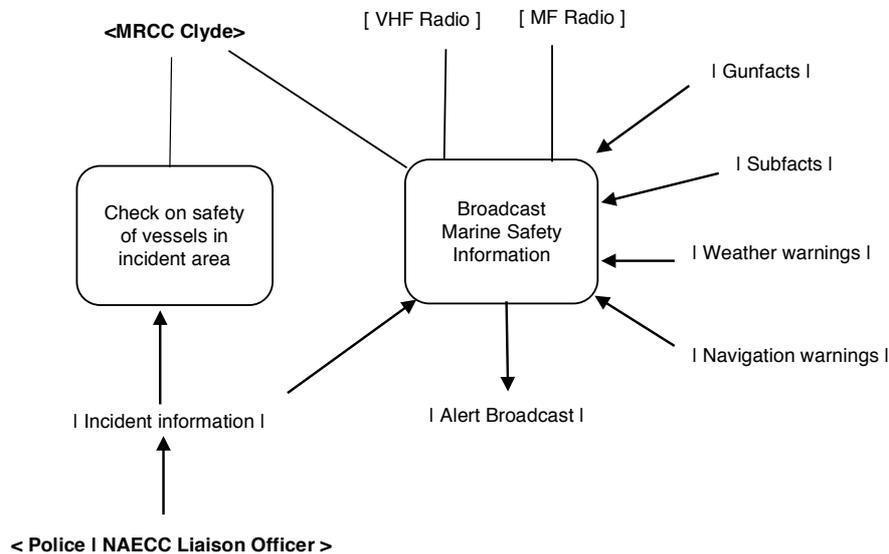

**Figure 3.** An example of a responsibility model

The broadcasting of safety information relies on a number of information resources from various sources (not shown here) and the physical resources of VHF and MF radios, which are used to broadcast the information.

## 5  Deriving information requirements

Responsibility models document the responsibilities of the agents involved in a multi-agency virtual enterprise and so serve as a basis for identifying the sources of requirements and the stakeholders who need to be consulted to derive these requirements. Our approach to requirements derivation is based on a set of structured questions that are put to stakeholders in the system. These questions are based around the following topics:

1. *What information needs to be provided to discharge this responsibility?*
   Whilst an apparently simple question, it is not necessarily the case that stakeholders from different agencies require the same information. For example, a stakeholder in agency A may already have some information because it is generated in agency A but this needs to be provided in other agencies. So, as well as identifying specific information items, these questions identify information that may have to be shared between agencies.
2. *What channels are used to communicate this information?*
   This question identifies the ways in which information is communicated to stakeholders. In some organizational systems, this is simple and straightforward but in other circumstances such as emergency response,

communication channels can be unreliable. We therefore may identify
   requirements for alternative communication channels that may be used.
3. *Where does this information come from?*
   Again, an apparently simple question that can elicit surprisingly complex answers. Our aim is to identify the databases and data sources for the information required but different stakeholders may actually acquire comparable information from different sources. The question can often reveal duplication and overlap in the information maintained by organizations.
4. *What information is generated and recorded in the discharge of this responsibility and why?*
   This question tries to tease out what information is created by an agent who holds and responsibility and the rationale for the information creation. This helps us identify requirements for storing that information and for maintaining meta-data for that information (who created, when created, etc.
5. *What channels are used to communicate this recorded information?*
   As above, we are interested in communication problems that may arise and backup requirements.
6. *What are the consequences if the information required is unavailable, inaccurate, incomplete, late, early?*
   Problems of information availability are common in multi-organizational systems Here we are specifically interested in trying to derive 'coping' requirements which allows the system to continue in effective operation when things go wrong. We have developed an approach based on HAZOPs [7], which we have discussed in some detail in a separate paper [8].

These questions are not formulaic – they are interpreted by the requirements engineering depending on local circumstances and the people being interviewed. Their purpose is to structure the discussion between a requirements engineer and system stakeholders. Typically, they lead to further questions and discussion about how stakeholders discharge their responsibilities. We expect the requirements engineer to deliver the results of that discussion in a form that is appropriate for the type of system being developed. This could be natural language requirements, diagrams or tables or even user stories.

## 6  Case study – emergency coordination system

To illustrate the derivation of requirements from responsibilities, we use an example of a system that helps coordinate the responses of the different agencies that are involved in dealing with civil emergencies. In the UK (and we understand elsewhere), the emergency services each have their own command and control systems and they do not think that it is appropriate to integrate these into a single system for all of the emergency services. Systems from other agencies may also be required to support emergency coordination. These might include systems from government agencies, such as the environment agency for flood management, and systems from local and regional government that maintain information about the local area.

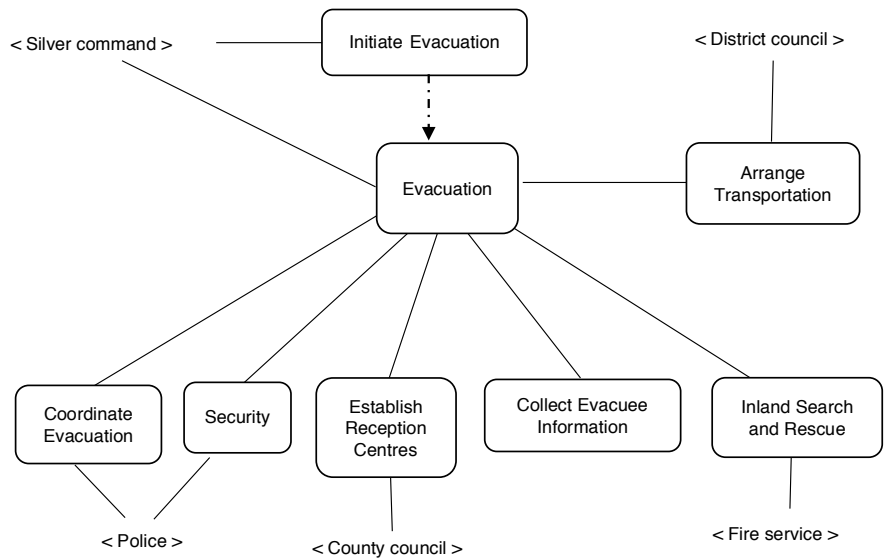

**Figure 4**. Responsibility model of evacuation coordination

Therefore, the coordination system is primarily an information management system that draws information from other systems and databases. It serves a variety of different stakeholders - emergency service staff working at the site of the emergency, emergency service coordination and planning officers, press officers, local government officers, and so on.

We will focus in this case study on the information requirements for the evacuation of premises in an area that is threatened by flooding. The information here is drawn from an analysis of a flooding emergency in the north-west of England in 2005.

Figure 4 shows the responsibility model for area evacuation. Some terminology here may need to be explained:
1. Silver command is the command centre that is set up to deal with the emergency and is responsible for strategic decision making. It is located in a pre-allocated, networked control room. Officers from the different services are involved in Silver Command. It is generally located away from the source of the emergency and communicates by radio and telephone with the on-site command centre (Bronze Command).
2. In England, there are two levels of local government at the district level (District Council) and at the regional level (County Council). The allocation of functions to District and County councils is historical.

Given situation information such as the current and predicted level of local rivers and weather forecasts, Silver Command carries out a risk analysis and on the basis of that analysis may decide that an area should be evacuated (Initiate Evacuation). This is a legal decision that results in the handover of certain powers to the emergency services, such as the right to remove people from their homes, and this must be agreed by all of the services. Evacuation then proceeds (the dashed arrow in Figure 4 means that these responsibilities are discharged in sequence). The police are responsible for coordination and the maintenance of security and the fire service are responsible for search and rescue operations if these are required. The district council arranges transport for evacuees but the county council is responsible for setting up safe places (reception centres) to which the evacuees are taken.

Notice that the responsibility to collect information about evacuees has no agent associated with it. This is an omission in the emergency plan that we discovered when we drew up the responsibility model. This highlights one of the key benefits of responsibility modeling – it serves to expose responsibility vulnerabilities that may lead to a failure to discharge the responsibility.

Figure 4 also illustrates another characteristic of responsibility models – they may be incomplete. In this case, we do not show any resources that may be used in the discharge of a responsibility. This means that we do not need to clutter a diagram with unnecessary information before using that diagram and that we can proceed with modeling even when information is incomplete.

By asking the questions identified above, we can discover the information that is required and produced as part of evacuation coordination. This is presented in a tabular form in Tables 1 and 2. Table 1 documents the information that is required to discharge the responsibility. Table 2 documents the information generated.

| **Information required** | **Source** | **Communication channel** |
|---|---|---|
| Area map | County council | Radio data link to printers in local command centre |
| Priority premises list | District Council | Radio data link to printers in local command centre |
| Assembly points list | District Council | Radio data link to printers in local command centre |
| Evacuated premises | Police, Fire Service | Radio from Silver Command |
| Unsafe routes | Police | Radio from Silver Command |
| Threat information | Environment agency | Radio from Silver Command |
| Transport capacity and availability | District Council | Radio from Silver Command |
| Police and other emergency service availability | Police, other services | Radio from Silver Command |

**Table 1.** Information used in the discharge of the evacuation responsibility

The priority premises list is a list of premises, such as schools and care homes, where the occupants cannot be expected to evacuate themselves. The evacuation involves local residents going to local assembly points from which they are transported to a place of safety. Unsafe routes are those routes that must be closed off by the emergency services because they are already flooded or in imminent danger of flooding.

We have found that it is important to maintain information about the communication channels that are used. Communications are often a problem in emergency management so it is important to check that backup channels are available. In addition, the system being developed automatically generates and sends messages and so it is important to have information about how these should be transmitted.

| Information created/recorded | Channels |
|---|---|
| Information about evacuated premises, evacuation time and units responsible for evacuation | Radio or verbal report from ground units to local Bronze Command. Email or fax to Silver Command if available, otherwise radio. |
| Information about unchecked premises | Radio or verbal report from ground units to local Bronze Command. Email or fax to Silver Command if available, otherwise radio |
| Information about unsafe routes | Radio or verbal report from ground units to local Bronze Command. Email or fax to Silver Command if available, otherwise radio |

**Table 2.** Information recorded in the discharge of the evacuation responsibility

A critical part of the questioning process is the analysis of the consequences if information is not available as expected. We assess these consequences when the information required is unavailable, inaccurate, incomplete, delivered late or early? For example, consider the information relating to the list of priority premises to be evacuated:

1. *Information unavailable*. A manual premises check is required to see if there are vulnerable people who need help with evacuation. Evacuation delayed and additional effort required.
2. *Information inaccurate*. Again, a manual premises check may be required. There may be delay in evacuating vulnerable people and vulnerable people may not left behind.
3. *Information incomplete*. Delay in evacuation.
4. *Information late*. Information has to be communicated to units in the field by radio rather than to local coordination centre. This is time consuming and less reliable than written communications with Bronze Command.
5. *Information early*. No consequence.

The information on 'information hazards' may then be used as a basis for defining requirements for mitigation strategies that lessen the consequences of subsequent failure. We see examples of these in the requirements shown in the following section.

### 6.1 System requirements

After the information about the information used by and generated by stakeholders has been collected, it is then the responsibility of the requirements engineer to generate system requirements in an appropriate form. If a formal requirements document is to be produced, this is likely to be a mix of natural language requirements and tables; if the requirements are expressed less formally, then tables such as Table 1 and Table 2, along with relevant commentary may be all that is required.

We show a subset of natural language requirements for an emergency response coordination system (ERCS) along with the rationale for these requirements below. These have been derived from the information documented in Tables 1 and 2.

1. The ERCS shall be able to import information from the District Council planning system, the Police emergency system and the Fire Service emergency system. (*Different types of information needs to be shared and this allows for information transfer between agencies*).
2. All information to be imported shall be available in either XML format or in PDF. (*This is intended to minimize the problems of importing information from different databases*).
3. The ERCS shall maintain its own list of priority premises to be evacuated for each town in the local area. This shall be updated by the local council when the coordination centre is established from the council's list. (*This is a critical asset for evacuation. The premises list is normally maintained by the local government authority but may not be immediately available outside of normal working hours; While an older list may be out of date, it is better than nothing*).
4. The ERCS shall maintain a list of premises evacuated along with the time of evacuation and the units involved in the evacuation. (*This allows units involved in the evacuation to be coordinated and maintains an audit trail of who did what and when*).
5. The ERCS shall notify all liaison officers of new information about the threat situation as it becomes available. (Different services may respond differently to changes in the threat situation e.g. local government staff may withdraw from a situation because they are not equipped to deal with search and rescue).
6. Alerts that threat information has changed should be displayed on all user screens and should be sent by SMS to all liaison officers (*Threat information is critical and should be sent on multiple channels. SMS can reach officers when they are not at their desk*).
7. ERCS operators should be able to update the Area Map with information about unsafe routes, without the need to access the source data for that map (*This allows maps to be distributed to emergency services but does not require operators to have access to the Council GIS*).
8. If information on evacuated premises is not available, the ERCS shall request the information from the Police liaison officer and send an SMS alert that this information has been requested. (*The Police are responsible for collecting this information and the Police liaison officer is then responsible for initiating a manual premises check if this is required*).
9. The ERCS shall maintain a list of all unchecked premises and shall automatically update this when information on evacuated premises is updated.

(*If premises have been evacuated, they are no longer unchecked. This partially mitigates problems due to delays in updating the unchecked premises list*).
10. Transcripts of all incoming radio communications shall be maintained in the ERCS along with the time of these communications and the identifier of the source of the message (*This is required for auditing purposes if problems are subsequently reported*).

## 7 Related work

The notion of using models of responsibility to support the requirements engineering process was first suggested by Dobson and Strens [8]. This was part of the ORDIT project [9, 10], which focused on organizational issues in software engineering. The work on requirements here was mostly concerned with what they termed 'organizational requirements' – requirements that are derived from organizational factors such as the power and authority relationships between people and departments in an organization.

Working in conjunction with Sommerville and others [11, 12], Dobson continued the work on responsibility models and documented this in a series of papers, which were published in a book that he co-edited with Dewsbury [13]. These were the basis for our own work on responsibility modeling where we have been concerned with responsibilities and system dependability and models of responsibility in virtual organizations [14, 15, 16] .

Responsibilities are an example of an abstraction that is clearly located in the world of system stakeholders rather than a technical abstraction such as an object or system function. The most closely related alternative abstraction to responsibility that has been proposed is the notion of a *goal*. A goal is seen as something that an agent is trying to achieve and goal-based approaches to requirements engineering such as i* and KAOS are intended to expose high level dependencies between the goals associated with agents in a given system [17, 18, 19].

Sub-goals may be derived from higher level objectives and assigned to agents for completion. Goals are achieved through the achievement of some or all sub- goals. Relationships between sub-goals express the possible ways in which the super-goal may be achieved. Analysis of such models can examine, for example, whether a super-goal may fail due to the failure of a single sub-goal (brittleness), or whether a particular agent has been overloaded with too many
goals to achieve.

We argue that the key benefit of using responsibilities rather than goals comes from the naturalness of the abstraction. Goals, in the sense of something that is to be achieved, have 3 main problems:
1. The goals of individuals are usually internalised and people find these very difficult to articulate. This is particularly true in professional roles where the work to be done is left to the discretion of the individual.
2. Many, perhaps most organizations, do not have a coherent set of organizational goals and, where they do, it cannot be assumed that goals set by management are actually shared by the people in the organization.

3. The goals of individuals in an organization may be focused on personal advancement and this may, in fact, conflict with organizational goals.

In a review of research on goal-oriented approaches, Lapouchnian [20] rightly states "Identifying goals is not an easy task". He has found, in practice, that goals are normally derived from other information that is discovered from stakeholders rather than articulated directly from them.

## 8  Conclusions

The modeling approach proposed here, based on the responsibilities that have been assigned to agents in an enterprise, has been found to be useful in supporting the elicitation of 'information requirements'. We argue that for enterprise systems, which are systems of systems it is more appropriate to focus on the information required and created by system stakeholders rather than the behavioural characteristics of a system.

The key benefits of using responsibilities and responsibility models in this context are:
1. N*aturalness: can stakeholders without experience of requirements engineering relate to the approach?* The notion of responsibility and responsible behaviour is widely used in everyday discourse so people can readily discuss their responsibilities in some situation. The questions used to discover information requirements relate directly to the stakeholder's job and are therefore easy to understand.
2. *Scalability: Can the approach be used with real rather than simple example systems?* The problem with many RE methods is that they have been developed using relatively simple systems and when scaled up, unmanageable volumes of information are created. Our development has always relied on real system examples and we are confident that our approach scales – for example, we have developed responsibility models of 300-page emergency plans.
3. *Complementarity: can the approach be used alongside other requirements engineering methods?* Responsibility models offer a different perspective from the behavioural perspective used in other methods so there are no problems in practice in using these together.

There are practical and methodological problems in attempting to compare requirements engineering methods, which mean quantitative comparative evaluation is unreliable. Furthermore, comparison of methods is not the same as comparison of outcomes. Method A may be better than method B at eliciting requirements but until a system has been implemented and put into use, we really don't know if these requirements meet the needs of the system stakeholders.

Therefore, we cannot and do not claim here that the use of responsibility models in the RE process necessarily leads to the discovery of 'better' requirements. All we can say is that responsibilities are a good way of stimulating requirements discussions and this, we believe, increases the chances that the requirements are likely to be appropriate.

Responsibility models provide a technology-independent perspective on complex systems of systems, where the components are already in existence. We have explored how these models may also be used in the systems design phase [21]. In this work, we have found the need to enhance these models with the notion of a capability – a set of competences and resources – that defines the responsibilities that could be assigned to a system. This work is still at an early stage but it points the way to how responsibilities and capabilities could be used to support system of systems design.